\documentclass[sigconf]{acmart}
\usepackage{graphicx}
\usepackage{multirow}

\usepackage{amssymb}
\usepackage{array}
\usepackage{url}
\usepackage{booktabs}

\AtBeginDocument{%
  \providecommand\BibTeX{{%
    \normalfont B\kern-0.5em{\scshape i\kern-0.25em b}\kern-0.8em\TeX}}}

\copyrightyear{2021}
\acmYear{2021}
\setcopyright{acmcopyright}\acmConference[ICMR '21]{Proceedings of the 2021 International Conference on Multimedia Retrieval}{August 21--24, 2021}{Taipei, Taiwan}
\acmBooktitle{Proceedings of the 2021 International Conference on Multimedia Retrieval (ICMR '21), August 21--24, 2021, Taipei, Taiwan}
\acmPrice{15.00}
\acmDOI{10.1145/3460426.3463584}
\acmISBN{978-1-4503-8463-6/21/08}

\begin{document}

\title{Cross-Modal Self-Attention with Multi-Task Pre-Training for Medical Visual Question Answering}

\author{Haifan Gong}
\authornote{Both authors contributed equally to this work. Guanbin Li is the corresponding author.}
\orcid{0000-0002-2749-6830}
\affiliation{%
  \institution{Sun Yat-sen University}
  \country{}
}
\email{gonghf@mail2.sysu.edu.cn}

\author{Guanqi Chen}
\authornotemark[1]
\affiliation{%
  \institution{Sun Yat-sen University}
  \country{}
}
\email{chengq26@mail2.sysu.edu.cn}

\author{Sishuo Liu}
\affiliation{%
  \institution{The University of Hong Kong}
  \country{}}
\email{sishuo@hku.hk}

\author{Yizhou Yu}
\affiliation{%
  \institution{The University of Hong Kong}
  \country{}}
\email{yizhouy@acm.org}

\author{Guanbin Li}
\authornotemark[2]
\affiliation{%
  \institution{Sun Yat-sen University}
  \country{}
}
\email{liguanbin@mail.sysu.edu.cn}


\begin{abstract}
Due to the severe lack of labeled data, existing methods of medical visual question answering  usually rely on transfer learning to obtain effective image feature representation and use cross-modal fusion of visual and linguistic features to achieve question-related answer prediction. These two phases are performed independently and without considering the compatibility and applicability of the pre-trained features for cross-modal fusion. Thus, we reformulate image feature pre-training as a multi-task learning paradigm and witness its extraordinary superiority, forcing it to take into account the applicability of features for the specific image comprehension task. Furthermore, we introduce a cross-modal self-attention~(CMSA) module to selectively capture the long-range contextual relevance for more effective fusion of visual and linguistic features. Experimental results demonstrate that the proposed method  outperforms existing state-of-the-art methods. Our code and models are available at \url{https://github.com/haifangong/CMSA-MTPT-4-MedicalVQA}.

\end{abstract}




\maketitle

\section{Introduction}
The medical visual question answering (VQA) aims to answer the questions of images in the medical domain. The common setting of medical VQA is to retrieval the answer from the answer set which best fits the given question and the image. With the expectation that VQA systems can not only provide clinicians with clinical decision support, but also help patients better understand their conditions based on medical images, several medical VQA datasets\cite{abacha2018nlm,lau2018dataset,abacha2019vqa} have been proposed. Since the questions and related answers are automatically generated, the medical VQA datasets produced by \cite{abacha2018nlm,abacha2019vqa} inevitably contain noise, which may not be the optimal choice for clinical decision support. Different from \cite{abacha2018nlm,abacha2019vqa}, the VQA-RAD dataset\cite{lau2018dataset} was manually constructed where clinicians naturally create the question-answer pairs about the radiology images. Unfortunately, there are only 315 images in the VQA-RAD dataset, while the out-performing deep learning based  algorithms require large-scale data for training. Therefore, many works\cite{lau2018dataset,nguyen2019overcoming,yan2019zhejiang,allaouzi2019encoder} adopt transfer learning to solve this problem, relying on external data for pre-training of image features before the training of feature fusion and answer prediction. However, those works neglect considerations for the compatibility and applicability of the pre-trained features for cross-modal fusion. In addition, compared to the VQA task of natural images, there are several unique challenges in the medical VQA domain, including semantic parsing of medical terminology, more complex cross-modal semantic alignment and fusion due to low contrast of medical images, and the multi-modal characteristics of medical images (i.e., CT, MRI, X-Ray).\par

\begin{figure*}[tbp]
\includegraphics[width=\textwidth]{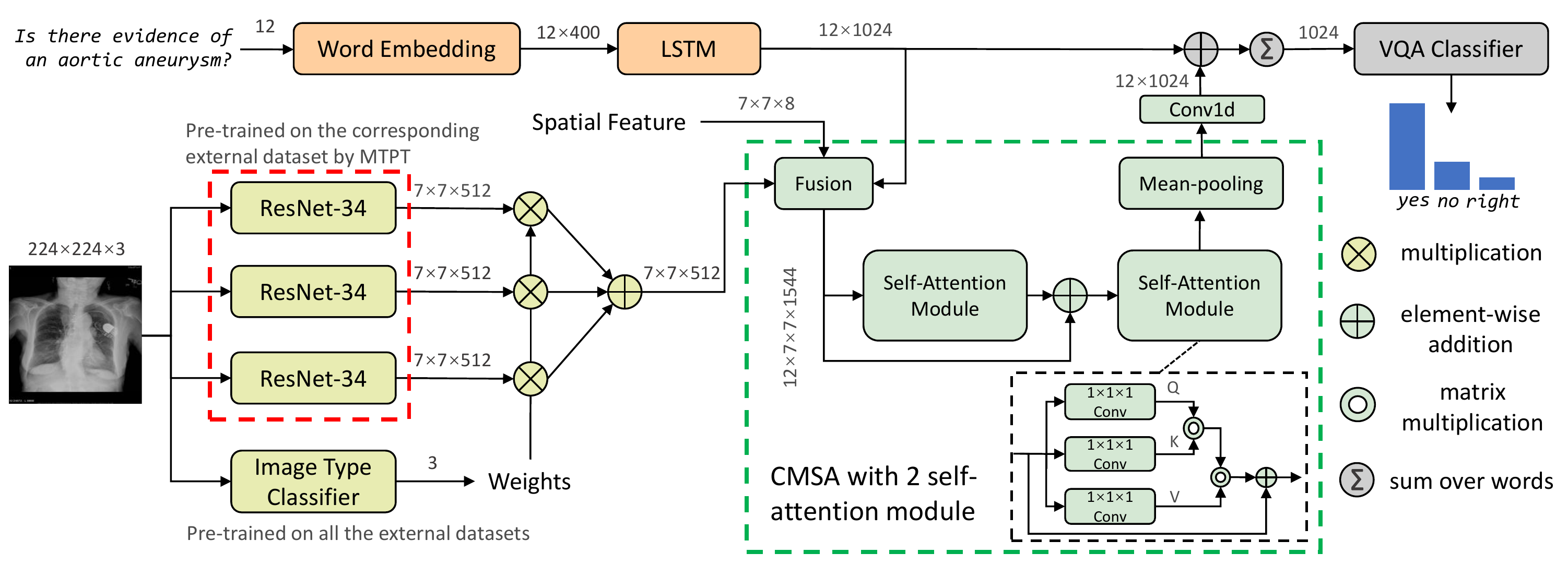}
\caption{Overview of the proposed medical VQA model. Our method consists of four components (with different colors in the figure): image feature extractor, question encoder, cross-modal self-attention (CMSA) module, and answer predictor.} \label{fig1}
\end{figure*}

Based on the above concerns, we propose to reformulate image feature pre-training as a multi-task learning paradigm, forcing it to take into account the applicability of features for both the specific image comprehension task and our proposed cross-modal fusion module. This has been proven to make more effective use of external data to better overcome the problem of data scarcity in medical VQA.  Secondly, a tailor-designed cross-modal self-attention (CMSA) module is used to effectively fuse the visual and linguistic features by learning and leveraging their long-range contextual relevance, which  effectively compensates for the low contrast and weak local feature representation in medical images through contextual information enhancement and complementation. Last but not least, we achieve the state-of-the-art performance on the VQA-RAD dataset.

\section{Related Work}
\subsection{Visual Question Answering}
With the prosperity of deep learning, VQA has received extensive attention in recent years and has made great progress, mainly benefiting from the strong bottom-up feature representation based on deep CNNs\cite{anderson2018bottom} and the cross-modal feature alignment and fusion techniques\cite{yang2016stacked,kim2018bilinear,yu2017multi,fukui2016multimodal,anderson2018bottom}. Anderson et al.  \cite{anderson2018bottom} proposed a bottom-up mechanism implemented by Faster R-CNN\cite{ren2015faster} to extract object-level representation for the input image, which achieved great success in both VQA and image captioning. From the perspective of cross-modal feature fusion, methods can be roughly divided into two main categories, including the attention based methods and multi-modal joint embedding. Anderson et al.\cite{anderson2018bottom} and Yang et al.\cite{yang2016stacked} developed different attention modules to adaptively attend on the relevant image regions based on the question representation. Kim et al. \cite{kim2018bilinear,yu2017multi,fukui2016multimodal} proposed to employ the compact bilinear pooling methods to combine the visual and linguistic features.

For medical VQA, the current common methods \cite{lau2018dataset,nguyen2019overcoming,yan2019zhejiang,vu2019ensemble,Jung2020vqa-med,vu2020question,ren2019cgmvqa} are to use CNN for image feature representation and leverage LSTM \cite{hochreiter1997long} or transformer-based methods (e.g., Bert \cite{devlin2019bert}, BioBert \cite{lee2020biobert}) to extract features for the given question. Varieties of general cross-modal fusion strategies (e.g., SAN\cite{yang2016stacked}, BAN\cite{kim2018bilinear}, and MFB\cite{yu2017multi}) are applied for feature fusion followed by the ultimate answer prediction. Compared to general VQA, medical VQA systems are required to comprehend medical terminology and focus on the corresponding visual content in the image. However, existing medical VQA methods do not realize the significance of these problems and directly borrow the general VQA technologies, which caused bottlenecks in the prediction accuracy of the models.

\subsection{Transfer Learning}
Due to the limitation of medical VQA data, many works rely on transfer learning to obtain effective image feature representation. In \cite{lau2018dataset,yan2019zhejiang,vu2019ensemble,chen2020hcp-mic}, they use a CNN which is pre-trained on ImageNet\cite{russakovsky2015imagenet} to encode medical image, such as VGGNet and ResNet. Allaouzi \textit{et al.} \cite{allaouzi2019encoder} utilizes CheXpert \cite{irvin2019chexpert}, a large dataset of chest radiographs, to pre-train a DenseNet-121 as the visual feature encoder. Nguyen \textit{et al.} \cite{nguyen2019overcoming} leverages a large scale of unlabeled medical images to pre-train its unsupervised denoising auto-encoder via a reconstruction task. However, there are no existing works attempt to consider the compatibility and applicability of the pre-trained features for cross-modal fusion, which is the emphasis of VQA models. 

\section{Methodology}
The proposed medical VQA framework is shown in Figure~\ref{fig1}, which includes a multi-task pre-training paradigm for more effective medical image representation learning, a cross-modal self-attention module for feature fusion, and the ultimate VQA classifier for question-related answer prediction. In this section, we elaborate the proposed multi-task pre-training method and the medical VQA model.

\begin{figure}[tbp]
\includegraphics[width=\linewidth]{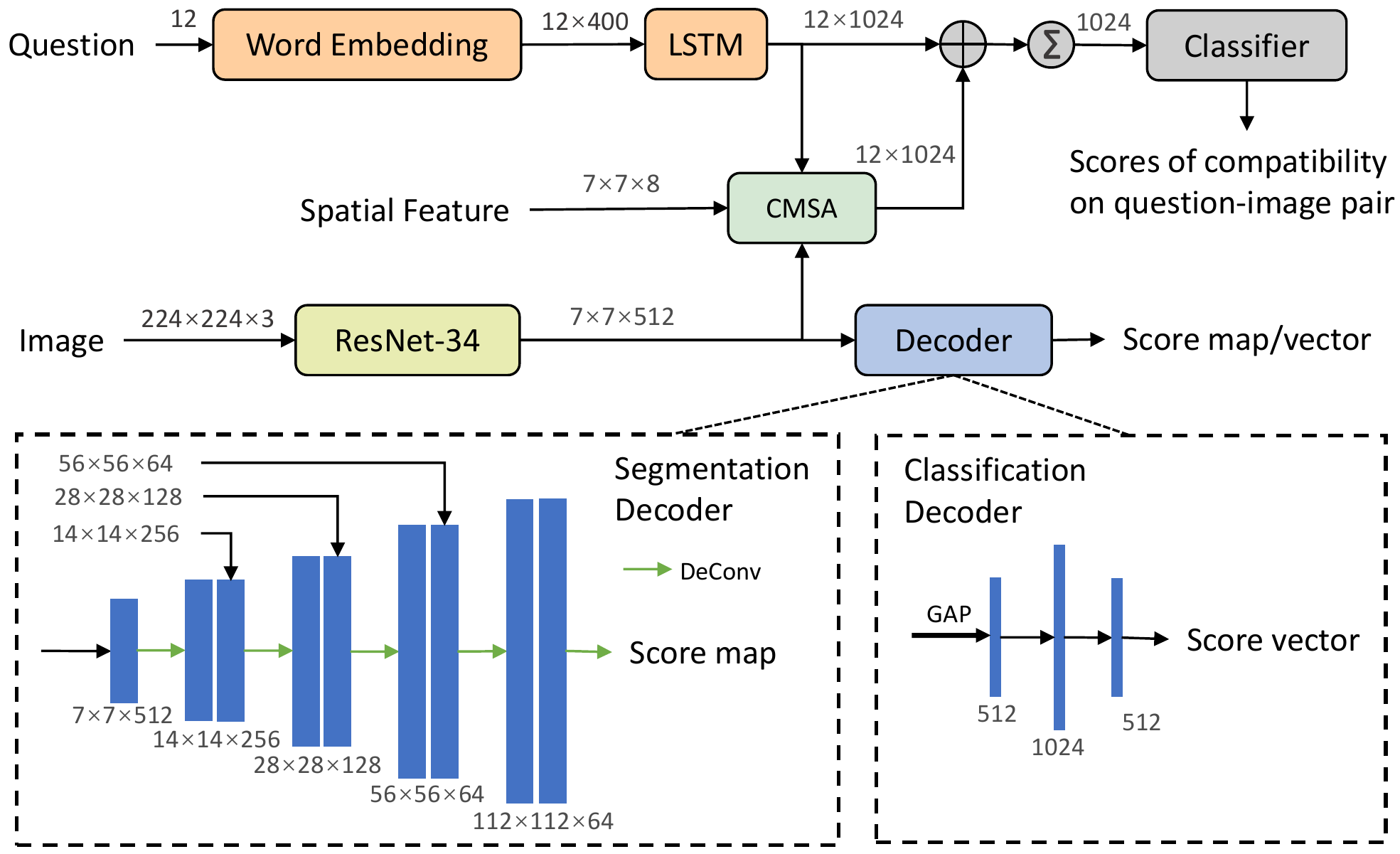}
\caption{Multi-Task Pre-Training: the model is jointly trained with an image understanding task and a question-image compatibility task. Depending on the dataset-specific image understanding task, the decoder can be selected as a fully convolutional network or a fully connected network.} \label{fig2}
\end{figure}

\subsection{Multi-Task Pre-Training}
During multi-task pre-training, our model is jointly trained with two separate tasks, including a regular image understanding task and a tailor-designed task for question-image compatibility test. The latter is defined as a binary classification task, which requires the model to determine whether the question is related to and suitable for a given image. For example, the question ``Are the lungs normal size?'' is suitable for a chest image rather than an abdominal image. For a given image from the external dataset, we randomly select a question from the VQA-RAD to form a question-image pair. The label for question-image compatibility testing is constructed by querying whether there exists a pair of the selected question and an image whose type is the same as the given image in VQA-RAD.\par
As shown in Figure~\ref{fig2}, during pre-training with external datasets, we use ResNet-34 \cite{he2016deep} as a backbone to capture the visual feature of the input image, and a decoder of symmetric structure for segmentation and a 3-layer MLP for image classification. For question-image compatibility testing, we use the proposed cross-modal self-attention (CMSA) module for feature fusion, which will be detailed later. It is worth noting that the CMSA module used here only contains one self-attention module without repeating, because we want this pre-training task focus on the representation learning of the image encoders rather than feature fusion. Formally, the multi-task loss function is defined as:
\begin{equation}
L=L_{spe}+L_{com}
\end{equation}
where $L_{spe}$ and $L_{com}$ are the cross-entropy loss for the specific image understanding task and the question-image compatibility task, respectively.

\subsection{Our Medical VQA Model}
Our medical VQA model consists of four parts: image encoding for capturing visual features of the given medical image, question encoding for extracting language features of the given question, cross-modal self-attention module for visual-language feature fusion, and answer prediction. We use a multi-task loss $L$ to train the proposed medical VQA model in an end-to-end fashion:
\begin{equation}
L=L_{vqa}+\alpha L_{type}
\end{equation}
where $L_{vqa}$ and $L_{type}$ are the cross-entropy loss for classification based answer prediction and image type classification. $\alpha$ is a hyperparameter for balancing the  two loss terms, which is set to 0.5.
\subsubsection{Image encoding:} 
Since clinicians use different imaging techniques to precisely diagnose the diseases of different organs, we use three separate ResNet-34 networks pre-trained on the corresponding external datasets to capture visual features of MRI, CT, X-Ray images, respectively. Then we use a classifier to determine the type of medical image and select the corresponding visual feature in a soft manner:
\begin{equation}
v = w_1 v_a + w_2 v_h + w_3 v_c
\end{equation}
where $v$ denotes the final visual feature, $v_a$, $v_h$ and $v_c$ denote the output features from the encoder corresponding to the abdomen, head and chest images respectively.  $w$ is the output vector of image type classifier with $\sum_{l=1}^{3}w_{l}=1$, which represents the corresponding weight of each medical image type. Besides, to better comprehend and answer questions related to the localization of local images, we follow \cite{ye2019cross} to obtain an 8-D spatial feature map $s$ with the same resolution as the visual feature $v$. The spatial vector at each position in the spatial feature map $s$ encodes the normalized coordinates of top-left, center, bottom-right, width and height of the grid.

\subsubsection{Question encoding:}
Following the previous work \cite{nguyen2019overcoming}, each input question is trimmed to a maximum of 12 words, and it is zero-padded when its length is less than 12. Each word is represented as a concatenation of a 200-D BioWordVec \cite{zhang2019biowordvec} word embedding and another 200-D augmenting embedding from the VQA-RAD. BioWordVec is a pre-trained biomedical word embedding based on PubMed and MeSH. Each 400-D word embedding vector is further fed into a LSTM to obtain the question embedding $q\in \mathbb{R}^{12 \times 1024}$.

\subsubsection{Cross-Modal Self-Attention:}
Before cross-modal fusion, we are given the visual feature $v \in \mathbb{R}^{7 \times 7 \times 512}$, spatial feature $s \in \mathbb{R}^{7 \times 7 \times 8}$ and question embedding $q\in \mathbb{R}^{12 \times 1024}$. For each word in the question, we concatenate its representation with visual and spatial features at each spatial location to produce a feature map $f \in \mathbb{R}^{7 \times 7 \times 1544}$. Then, we collect all the concatenated feature maps to obtain a multi-modal feature map $F \in \mathbb{R}^{12 \times 7 \times 7 \times 1544}$. Inspired by the effectiveness of self-attention\cite{wang2018non,vaswani2017attention} in capturing non-local context, we design our cross-modal alignment and fusion method. \par
Firstly, we linearly transform the multi-modal feature map $F$ to produce three feature maps $Q$, $K$, $V$ $\in \mathbb{R}^{12 \times 7 \times 7 \times 772}$ through three $1 \times 1 \times 1$ convolutional layers . We reshape them to the dimension $\mathbb{R}^{588 \times 772}$, and use  $Q$ and $K$ to compute the attention map $A$:
\begin{equation}
A = softmax(QK^T)
\end{equation}
where $A$ $\in \mathbb{R}^{588 \times 588}$ indicates the correlation of features in different positions. We multiply the attention map $A$ and the feature map $V$ to obtain the enhanced multi-modal representation $F^\prime \in \mathbb{R}^{588 \times 772}$:
\begin{equation}
F^\prime = A V.
\end{equation}
Next, we turn the dimension of $F^\prime$ to $\mathbb{R}^{12 \times 7 \times 7 \times 1544}$ through reshaping and $1 \times 1 \times 1$ convolution. The above operations are shown in the Figure~\ref{fig1} named self-attention module. Inspired by the `glimpse' in BAN\cite{kim2018bilinear}, we repeat the self-attention module again with residual connection. The final multi-modal representation $\hat{F} \in \mathbb{R}^{12 \times 1544}$ is obtained by applying a mean-pooling operation to the output of the residual connection between $F^\prime$ and $F$ over all spatial locations:
\begin{equation}
\hat{F}_i = \frac{\sum_{j=1}^{7} \sum_{k=1}^{7} \left(F^\prime_{i j k}+F_{i j k}\right)}{7 \times 7}
\end{equation}
where $i$, $j$, $k$ are the indices of the number of words, height and width of the feature map. Then $\hat{F}$ is transformed to the same dimension as the question embedding $q$ with a linear layer.
\subsubsection{Answer prediction:}
The joint representation $\hat{F}$ is added element-wise with question embedding $q$, and  it is summed over all words in the question. Finally, we feed it into a 2-layer MLP for answer prediction. Prediction score $s$ of the answer is calculated by:
\begin{equation}
s = MLP(\sum_{i=1}^{12} (\hat{F}_i + q_i)).
\end{equation}

\section{Experiments}
\subsection{Datasets and Metrics}
\subsubsection{Datasets:}
The proposed method is evaluated on the VQA-RAD dataset\cite{lau2018dataset}, which contains 315 radiological images with 3064 training questions and 451 test questions. We resort to three external  datasets to pre-train the visual encoders of different image types, including abdominal CT  \footnote{\url{https://www.synapse.org/\#!Synapse:syn3193805/wiki/217753}}, brain MRI\cite{Cheng2017} and chest X-Ray \footnote{\url{https://www.kaggle.com/paultimothymooney/chest-xray-pneumonia}}. The abdominal CT dataset includes 2178 images of  13 classes for multi-organ segmentation, where we use 2070 images for training and 108 images for validation. The brain MRI dataset comtains 3 types of brain tumors with 3604 images. We divide it into 3000 images and 64 images for training and validation, respectively. The chest X-Ray dataset contains 5232 images of `pneumonia' or `normal', which is split into 5000 images for training and 232 images for validation.

\subsubsection{Evaluation metrics:}
We use accuracy as the metric for the VQA task and the classification task during pre-training, where $acc_{cls}$ and $acc_{com}$ refer to the accuracy of the  image classification task and the question-image compatibility task respectively. Mean Intersection-over-Union (mIoU) is the criteria for segmentation.

\begin{table}[]
\caption{Comparisons with the state-of-the-art methods on the VQA-RAD test set. $para^\star$ means using not only the "freeform" but also the "para" answer type in the test set.}
\label{tab1}
\begin{tabular}{@{}cccc@{}}
\toprule
Methods              & Open   & Closed & All    \\ \midrule
SAN-RAD {[}10{]}     & 24.2\% & 57.2\% & 44.0\% \\
MCB-RAD {[}10{]}     & 25.4\% & 60.6\% & 46.5\% \\
SAN-MEVF {[}11{]}    & 40.7\% & 74.1\% & 60.8\% \\
BAN-MEVF {[}11{]}    & 43.9\% & 75.1\% & 62.6\% \\
Ours                 & 56.1\% & 77.3\% & 68.8\% \\ \midrule
BAN-CR-para$^\star$ {[}22{]} & 60.0\% & 79.3\% & 71.6\% \\
Ours-para$^\star$            & 61.5\% & 80.9\% & 73.2\% \\ \bottomrule
\end{tabular}
\end{table}

\begin{table}[]
\caption{Ablation study on the 3 external validation set.}\label{tab2}
\begin{tabular}{@{}ccccccc@{}}
\toprule
           & \multicolumn{2}{c}{Abdominal CT} & \multicolumn{2}{c}{Brain MRI} & \multicolumn{2}{c}{Chest X-Ray} \\ 
Methods    & mIOU           & $acc_{com}$    & $acc_{cls}$  & $acc_{com}$  & $acc_{cls}$   & $acc_{com}$   \\ \midrule
Baseline   & 0.682          & -              & 98.4\%       & -            & 97.8\%        & -             \\
MTPT    & 0.710          & 78.7\%         & 98.4\%       & 89.1\%       & 98.7\%        & 83.6\%        \\ \bottomrule
\end{tabular}
\end{table}

\subsection{Comparison with the state-of-the-art}
As shown in Table~\ref{tab1}, our proposed method is compared with 5 existing state-of-the-art approaches\cite{lau2018dataset,nguyen2019overcoming, Zhan2020}, and achieves the highest accuracy on both open-ended and closed-ended VQA. Compared with the advanced approach BAN-MEVF\cite{nguyen2019overcoming} using external datasets, the proposed method outperforms it by 12.2\% and 2.2\% w.r.t. accuracy on open-ended and closed-ended VQA, respectively. For the sake of fairness, the comparison between the proposed and the current best model BAN-CR\cite{Zhan2020} is based on the "freeform" and "para" questions as the setting in their code. The proposed method outperforms BAN-CR\cite{Zhan2020} by 1.6\% w.r.t. accuracy on all the questions. Nevertheless, the proposed method could be combined with the conditional reasoning\cite{Zhan2020} to gain a further improvement.

\subsection{Ablation Study}
To explore the effectiveness of the multi-task pre-training method diagram, we compare it with a single-task pre-training method which only pre-train on the external datasets for the original image classification or segmentation task. In Table~\ref{tab2}, `Baseline' represents the single-task pre-training method. `MTPT' denotes the designed method with BioWordVec word embedding. The results show that our proposed multi-task pre-training method can slightly improve the performance of each specific image understanding task.\par

\begin{table}[]
\caption{Ablation study on the VQA-RAD test set.}\label{tab3}
\begin{tabular}{@{}cccc@{}}
\toprule
Methods         & Open & Closed & All    \\ \midrule
INPT-CMSA   & 30.9\%     & 73.5\%      & 56.5\% \\
STPT-CMSA   & 41.5\%     & 74.1\%      & 61.0\% \\
MTPT-BAN    & 56.1\%     & 75.7\%      & 67.9\% \\
MTPT-CMSA   & 56.1\%     & 77.3\%      & 68.8\% \\ \bottomrule
\end{tabular}
\end{table}

After pre-training the visual encoders, we load the pre-trained weights to train the whole VQA model on VQA-RAD. To clearly illustrate the ablation study in Table~\ref{tab3}, we give the definitions as: (1) `INPT' uses three pre-trained ResNet-34 on ImageNet as visual encoders. `STPT' initializes the visual encoders  with single-task pre-training while `MTPT' loads the weights of visual encoders from the multi-task pre-training. (2)`CMSA' uses the proposed `CMSA' for feature fusion while `BAN' applies BAN\cite{kim2018bilinear} for feature fusion.

From Table~\ref{tab3}, `MTPT-CMSA' outperforms `STPT-CMSA' significantly with the same external datasets for pre-training, which suggests that the pre-trained visual features from our multi-task learning paradigm are more suitable for our CMSA module to obtain an effective multi-modal representation. Furthermore, the proposed `CMSA' feature fusion surpasses the `BAN' feature fusion method by capturing the long-range contextual relevance.

\section{Conclusion}
This paper introduces a distinguished medical VQA framework which is based on multi-task pre-training paradigm for more effective medical image representation learning. Moreover, the proposed CMSA module effectively fusion of visual and language features by capturing the long-range contextual relevance. Experimental results verify that our proposed method can leverage external data more effectively to overcome the limitation of medical VQA data. In the future, we gonna focus on integrating domain knowledge into medical VQA on the recent knowledge based dataset \cite{liu2021slake} for the interpretable medical application.

\section*{Acknowledgments} 
This work was supported in part by the Guangdong Basic and Applied Basic Research Foundation (No.2020B1515020048) and in part by the National Natural Science Foundation of China (No.61976250 and No.U1811463). This work was also sponsored by CCF-Tencent Open Research Fund.
\bibliographystyle{ACM-Reference-Format}
\bibliography{ref}
\end{document}